\documentclass[11pt]{article}

\usepackage{amsmath} 
\usepackage{amsfonts} 
\usepackage{amssymb}             
\usepackage[dvips]{graphicx} 
\begin{document}

\newcommand{\sw}{\!\not \!}
\newcommand{\del}{\partial}
\newcommand{\noi}{\noindent}
\newcommand{\idd}{\mathbb{I}}
\newcommand{\nc}{\hat{n}_{i}}
\newcommand{\mc}{\hat{m}_{i}}
\newcommand{\mn}{\hat{\mbox{\boldmath$m$}}\cdot\hat{\mbox{\boldmath$n$}}}
\newcommand{\tg}{\mbox{tg}\,}
\newcommand{\qm}{\mbox{\boldmath$q$}}
\newcommand{\um}{m_{u}}
\newcommand{\cm}{m_{c}}
\newcommand{\tm}{m_{t}}
\newcommand{\dm}{m_{d}}
\newcommand{\sm}{m_{s}}
\newcommand{\bm}{m_{b}}
\newcommand{\elm}{\mbox{\boldmath$e$}}
\newcommand{\mum}{\mbox{\boldmath$\mu$}}
\newcommand{\tam}{\mbox{\boldmath$\tau$}}
\newcommand{\nelm}{\mbox{\boldmath$\nu_e$}}
\newcommand{\nmum}{\mbox{\boldmath$\nu_\mu$}}
\newcommand{\ntam}{\mbox{\boldmath$\nu_\tau$}}
\newcommand{\ru}{r_u}
\newcommand{\rd}{r_d}
\newcommand{\rl}{\mbox{\boldmath$r_{\ell}$}}
\newcommand{\ruun}{\mbox{\boldmath$r^u_1$}}
\newcommand{\rude}{\mbox{\boldmath$r^u_2$}}
\newcommand{\rdun}{\mbox{\boldmath$r^d_1$}}
\newcommand{\rdde}{\mbox{\boldmath$r^d_2$}}
\newcommand{\rlun}{\mbox{\boldmath$r^{\ell}_1$}}
\newcommand{\rlde}{\mbox{\boldmath$r^{\ell}_2$}}
\newcommand{\rnun}{\mbox{\boldmath$r^{\nu}_1$}}
\newcommand{\rnde}{\mbox{\boldmath$r^{\nu}_2$}}

%\newcommand{\PreserveBackslash}[1]{\let\temp=\\#1\let\\=\temp}
%\let\PBS=\PreserveBackslash
%\newlength{\tmplength}
%\newenvironment{TabularC}[1]{\end{tabular*}\par}
\newcommand{\Det}{\mbox{det}\,}
\newcommand{\Tr}{\mbox{tr}\,}

\title{
\vspace{-2.25cm}
\parbox{\textwidth}{\small
    \begin{center}
      \hfill
      \parbox{6cm}{
        \begin{flushright}
          UCL-IPT-01-06
        \end{flushright}
        }
    \end{center}
    }\\[0.5cm]
\huge Natural Relations in the Standard Model}

\date{\vspace{-1.5cm}}

\author{M. Buysse\thanks{E-mail: buysse@fyma.ucl.ac.be} \\
{\it Institut de Physique Th{\'e}orique }\\
{\it Universit{\'e} Catholique de Louvain}\\
{\it 2, Chemin du Cyclotron} \\
{\it B-1348 Louvain-La-Neuve, Belgium}\\
}

\maketitle

\begin{abstract}
We establish the potential existence of {\it natural relations}
between the Cabibbo angle and the quark mass ratios, in a Standard
Model with one Higgs doublet and two quark generations. The argument is
based on the calculation of the divergent one-loop radiative
corrections to the quark mass matrices.
\end{abstract}

\normalsize

\begin{equation*}
\end{equation*}

\vspace{-1cm}

One of the outstanding questions raised by the Standard
Model (SM) of elementary particles concerns the set of free parameters
it seems to involve. Many of those parameters find their origin in the
Yukawa sector of the theory: six
quark masses, three mixing angles and one phase, not to mention
the leptons. Furthermore, their empirical values show a strongly
hierarchical pattern. Since the early times of its discovery, one has tried to
complete the SM in order to reduce the size of its set of free parameters
and hence propose an explanation for this observed hierarchy. 
The simplest way to reach this goal is to look for potential relations
between apparently free parameters. But one has to make sure that
those relations are preserved by the renormalization; 
they must be {\it natural} \cite{Weinberg:1974, Weinberg:1977}. 
Until now, most attempts consisted in enlarging the symmetry
group of the SM by
adding a horizontal component to it 
\cite{Fritzsch:1977, Harari:1978, Pakvasa:1978, Fritzsch:1990}. 
The horizontal symmetry imposes constraints on the structure of the
Yukawa couplings. After spontaneous breakdown of the symmetry, the
fermion mass matrices that are generated still bear the stamp of those
constraints and through bidiagonalization, they give rise to relations
between mass ratios and mixing angles. Such an implementation
guarantees that the relations survive to renormalization; they are
called {\it natural}. 
However one soon realized \cite{Wyler:1979, Barbieri:1978,
  Gatto:1979a, Gatto:1979b} that it could not be achieved without
extending the particle content -- by considering models with more than
one Higgs doublet, thereby increasing
the number of couplings... 

We consider the SM in its minimal realization, namely built up with one
single Higgs doublet. We ignore the leptons and concentrate on
the quark sector. The approach we put forward here is based on the calculation of the
divergent one-loop radiative corrections to the quark mass matrices,
which exclusively involves self-energy and tadpole diagrams. More precisely, since we are
interested in natural relations between up-type quark mass ratios and 
down-type quark mass ratios on the one hand, and mixing angles on the
other hand, we will solely compute the divergent one-loop radiative
corrections to those specific parameters. This considerably simplifies
our task. One indeed notices that neither QED nor QCD, which are
flavour-blind, will bring in divergent contributions that would affect the
mixing angle or the mass ratios. The same argument holds for the diagrams involving the
transverse polarizations of the $Z^{0}$ and of the
$W^{\pm}$ vector bosons, as well as the tadpoles. 
To persuade oneself, it is worth checking that the contribution of the
latter diagrams to the renormalization of the quark mass ratios reads
\begin{equation*}
\frac{m_{u}}{m_{c}} \longmapsto
\frac{(1+C_{\gamma}+C_{G}+C_{Z^{0}}+C_{W^{\pm}}+C_{T})m_{u}}
{(1+C_{\gamma}+C_{G}+C_{Z^{0}}+C_{W^{\pm}}+C_{T})m_{c}} 
=\frac{m_{u}}{m_{c}} 
\end{equation*}

\noi where $C_{\gamma}$, $C_{G}$, $C_{Z^{0}}$ and $C_{W^{\pm}}$ 
respectively originate from the interventions of the photon, the
gluons, the transverse $Z^{0}$ and the transverse $W^{\pm}$ in the up-type quark
self-energies, while $C_{T}$ originates from the tadpole diagrams. 
Those $C$'s are identical for the mass renormalization of
any up-type quark -- as far as the divergent part is concerned. The
reasoning can be applied to the down-type quarks. For
the mixing angles, the proof is even more direct since none of those
diagrams leads to divergent non-diagonal correction to the tree-level
diagonal mass matrices. 
In other words, the only diagrams one has to consider are the quark self-energies
due to the exchange of the scalars (Higgs and would-be-Goldstone
bosons) -- for
the complete list of the relevant divergent diagrams, see figure
1. This is not astonishing since the scalars are the only fields
that know about the difference between the fermion families. We are
looking for a special structure inside the Yukawa couplings
(or the mass matrices), hence the fact that the Yukawa
sector is the only one responsible for the naturalness of this
structure will come as no surprise. 

We look into a two-quark-generation SM. Once we have the
expression of the divergent one-loop radiative corrections to the
Cabibbo angle and to the two mass ratios, we suppose the existence of a
tree-level natural relation between them. Since it is natural, this putative
relation must hold at the one-loop level up to some finite
corrections. This gives us a severe criterion to constrain the shape
of any natural relation. We show that there are indeed potential
one-loop natural relations in a SM with one Higgs doublet and two quark generations,
contrary to what has been thought so far. We
conclude that those
relations cannot originate from any additionnal horizontal symmetry. 
We finally state our a result in a different way and claim there is
one apparently free parameter, in a SM with one Higgs doublet and two quark generations,
to which the one-loop radiative corrections are finite. In other
words, one apparently free parameter whose presence in the SM
seems not to be necessary to render it renormalizable; or else, one apparently
free parameter whose one-loop $\beta$-function vanishes. We call it a {\it
  stable} parameter. 

\vspace{1cm}

Let us briefly recall the input we need. We start from a two-fermion-generation SM; the scalar
sector includes one single Higgs doublet. The hadronic Yukawa sector, which we are
interested in, contains three useful free parameters: the Cabibbo angle
$\theta$ \cite{Cabibbo:1963} and the two mass ratios $\ru =\um/\cm$ and  $\rd
=\dm/\sm$. The approach is the following: we assume the
existence of a natural relation between the Cabibbo angle and the mass
ratios; we then make use of the definition of a natural relation to
establish its general form. Namely, assuming the existence of
a natural relation automatically determines that relation. 

A general natural relation between the Cabibbo angle and the mass
ratios reads: 
\begin{equation}
F(\theta)=G(\ru , \rd)
\label{relnat}
\end{equation}

\noi so that 
\begin{equation}
F(\theta + \delta \, \theta) 
= G(\ru + \delta \, \ru , \rd + \delta \, \rd)
\label{propnat}
\end{equation}

\noi where $\delta \, \theta$, $\delta \, \ru $ and $\delta \, \rd$ are the divergent
parts of the radiative corrections to $\theta$, $\ru$ and $\rd$
respectively. This last expression is the mathematical translation of
naturalness. To be natural, (\ref{relnat}) cannot be broken by
infinite radiative corrections. That is, (\ref{propnat}) must be
verified. Thus
\begin{equation*}
F_{\shortmid \theta}\, \delta \, \theta 
=
G_{\shortmid r_{u}}\, \delta \, \ru 
+ 
G_{\shortmid r_{d}} \, \delta \, \rd 
\end{equation*}

\noi where $f_{\shortmid x}$ denotes the (partial) derivative of $f$
with respect to $x$. 
Computing the one-loop radiative corrections to the Yukawa couplings 
(see Appendix), one obtains: 
\begin{equation*}
\delta \, \theta 
 = 
\epsilon \left [
\frac{1+\ru ^{2}}{1-\ru ^{2}}(\dm ^{2} - \sm ^{2})
+ 
\frac{1+\rd ^{2}}{1-\rd ^{2}}(\um ^{2} - \cm ^{2})
\right ]
\sin \theta \cos \theta 
\label{vartheta}
\end{equation*}
\vspace{-4mm}
\begin{equation}
\delta \, \ru
=
\epsilon \; \ru \;
[ (\dm ^{2} - \sm ^{2})
\cos 2 \theta 
- (\um ^{2} - \cm ^{2})
]
\label{varpara}
\end{equation}
\vspace{-8mm}
\begin{equation*}
\delta \, \rd
= 
\epsilon \; \rd \;
[ (\um ^{2} - \cm ^{2})
\cos 2 \theta 
- (\dm ^{2} - \sm ^{2})
]
\end{equation*}

\noi with $\epsilon = \frac{3}{4} \frac{2}{v^{2}}\left (\frac{1}{4\pi^{2}}\ln
  \frac{\Lambda^{2}}{\mu^{2}} \right )$, $\Lambda$ being a cut-off, $\mu$ an
arbitrary energy scale and $v$ the vacuum expectation value of the
scalar field. Let us bear in mind that we do not care about the finite part
of the corrections. Introducing those expressions into equation (\ref{propnat})
splits it into two independent equations (since we are not interested in
considering relations involving mass ratios different from $\ru$ and
$\rd$): 
\begin{equation}
 F_{\shortmid \theta}\sin \theta \cos \theta 
\,\frac{1+\ru ^{2}}{1-\ru ^{2}}
= 
G_{\shortmid r_{u}}\,\ru \cos 2 \theta
-
G_{\shortmid r_{d}}\,\rd 
\label{sys1}
\end{equation}
\vspace{-4mm}
\begin{equation}
 F_{\shortmid \theta}\sin \theta \cos \theta 
\,\frac{1+\rd ^{2}}{1-\rd ^{2}}
= 
G_{\shortmid r_{d}}\,\rd \cos 2 \theta
- 
G_{\shortmid r_{u}}\,\ru 
\label{sys2}
\end{equation}

%\noi qui, choisissant $\ru\not = \rd$, ne sont compatibles que pour
%$\theta \in \{0,\frac{\pi}{4},\frac{\pi}{2}\}$.  En effet, par
%{\'e}limination de $F_{,\theta}(\theta)\sin \theta \cos \theta $ et
% de la d{\'e}pendance angulaire, on obtient les deux
%{\'e}quations 
%\begin{equation*}
%\frac{1+\rd ^{2}}{1-\rd ^{2}}\,
%G_{,r_{u}}(\ru , \rd)
%=
%\frac{1+\ru ^{2}}{1-\ru ^{2}}\,
%G_{,r_{d}}(\ru , \rd)
%\end{equation*}
%\vspace{-4mm}
%\begin{equation*}
%\frac{1+\rd ^{2}}{1-\rd ^{2}}\,
%G_{,r_{d}}(\ru , \rd)
%=
%\frac{1+\ru ^{2}}{1-\ru ^{2}}\,
%G_{,r_{u}}(\ru , \rd)
%\end{equation*}
\noi which turn out to be compatible if and only if 
\begin{equation}
\cos 2\theta 
= 
\frac
{
\frac{1-\ru ^{2}}{1+\ru ^{2}}\,
\rd\,
G_{\shortmid r_{d}}
-
\frac{1-\rd ^{2}}{1+\rd ^{2}}\,
\ru\,
G_{\shortmid r_{u}}
}
{
\frac{1-\ru ^{2}}{1+\ru ^{2}}\,
\ru\,
G_{\shortmid r_{u}}
-
\frac{1-\rd ^{2}}{1+\rd ^{2}}\,
\rd\,
G_{\shortmid r_{d}}
}
\label{identif}
\end{equation}

\noi Now this {\it must} be the natural relation
(\ref{relnat}) whose
existence has been supposed, i.e.\footnote{One checks that the
arbitrary character of those identifications will not
show itself in the expected solution. To prove it, we imagine
(\ref{eqangle}) would rather read $f(F(\theta))=\cos 2 \theta$. One should
then replace $G$ by $f(G)$ in the left-hand side of
(\ref{eqmasses}). But the right-hand side of it is
invariant under $G\mapsto f(G)$.  Namely, one can solve
(\ref{eqmasses}) with respect to the variable $f(G)$ which we finally
identify to $f(F(\theta))=\cos 2 \theta$.  }
\begin{equation}
F(\theta)=\cos 2 \theta
\label{eqangle}
\end{equation}

\noi and
\begin{equation}
%\mbox{\fbox{$
G(\ru ,\rd)
= 
\frac
{
%\textstyle
\frac{1-\ru ^{2}}{1+\ru ^{2}}\,
\rd\,
G_{\shortmid r_{d}}
-
\frac{1-\rd ^{2}}{1+\rd ^{2}}\,
\ru\,
G_{\shortmid r_{u}}
}
{
%\textstyle
\frac{1-\ru ^{2}}{1+\ru ^{2}}\,
\ru\,
G_{\shortmid r_{u}}
-
\frac{1-\rd ^{2}}{1+\rd ^{2}}\,
\rd\,
G_{\shortmid r_{d}}
}
%$}}
\label{eqmasses}
\end{equation}

\noi Exploiting the remaining information in (\ref{sys1}) and
(\ref{sys2}), and using (\ref{identif}) and (\ref{eqangle}),
yields 
\begin{equation*}
\begin{cases}
\frac{1+\ru ^{2}}{1-\ru ^{2}} G
- \ru G_{\shortmid r_{u}} 
\!\!\!\!\!\! & + \frac{1+\rd ^{2}}{1-\rd ^{2}} 
 = 0 
\\
\frac{1+\rd ^{2}}{1-\rd ^{2}} G
- \rd G_{\shortmid r_{d}} 
\!\!\!\!\!\! & + \frac{1+\ru ^{2}}{1-\ru ^{2}} 
 = 0
\end{cases}
\end{equation*}

\noi This system -- from which one obviously recovers equation (\ref{eqmasses})
-- is integrable, and the general solution reads
\begin{equation*}
G(\ru , \rd)
= \frac{ - (1+\ru ^{2})(1+\rd ^{2}) + 2 \lambda \: \ru \rd}
{(1-\ru ^{2})(1-\rd  ^{2})}
\end{equation*}

\noi where $\lambda$ is the integration constant. One concludes that, if a
relation of the kind suggested in (\ref{relnat}) exists, it necessarily belongs to
the following class:
%\begin{equation}
%%\mbox{\fbox{$
%\cos 2\theta
%= \frac{\textstyle - (1+ \frac{\um ^{2}}{\cm ^{2}})(1+\frac{\dm
%    ^{2}}{\sm ^{2}}) + 2\lambda \: \frac{\um}{\cm} \frac{\dm}{\sm}}
%{\textstyle (1-\frac{\um ^{2}}{\cm ^{2}})(1-\frac{\dm
%    ^{2}}{\sm ^{2}})}
%%$}}
%\label{sol}
%\end{equation}
\begin{equation}
%\mbox{\fbox{$
\cos 2\theta
= \frac{\textstyle - (\um ^{2} + \cm ^{2}) (\dm^{2} + \sm ^{2}) 
+ 2\lambda \: \um\cm\dm\sm}
{\textstyle (\um ^{2} - \cm ^{2}) (\sm ^{2} - \dm ^{2})}
%$}}
\label{sol}
\end{equation}

\noi We have found an infinite number of potential one-loop natural
relations inside the Yukawa sector. This set is parametrized by a
dimensionless constant $\lambda$. One has no further theoretical argument to
constrain the value of $\lambda$; and one cannot evade the difficulty by
asking one of the quark masses to vanish, since it would lead to a cosine
smaller than minus one\footnote{Consequently we may already conclude that it is not possible to naturally
set the mass of one single quark to zero, together with the
requirement of the existence of a natural relation between the Cabibbo
angle and the quark mass ratios. In this context, by excluding $\um = 0$ our result
rules out the natural vanishing of the QCD vacuum parameter
$\theta_{S}$.}. Now, one has to select the unique viable
natural relation by making $\lambda$ fit the data, i.e.
%\begin{equation}
%%\mbox{\fbox{$
%\lambda
%= \frac{\textstyle (1+\ru ^{2})(1+\rd ^{2}) + (1-\ru ^{2})(1-\rd ^{2})\cos 2\theta}
%{\textstyle 2 \ru \rd }
%%$}}
%\label{parlib}
%\end{equation}
%
\begin{equation}
%\mbox{\fbox{$
%\lambda
%= \frac{\textstyle \left (\um ^{2}\dm ^{2} + \cm ^{2}\sm ^{2}
%\right ) \cos ^{2} \theta
%+ \left ( \cm ^{2}\dm ^{2} + \um ^{2}\sm ^{2}
%\right ) \sin ^{2} \theta}
%{\textstyle \um\cm\dm\sm}
\lambda
= \frac{\textstyle (\um ^{2} + \cm ^{2}) (\dm^{2} + \sm ^{2}) + (\um ^{2} - \cm ^{2}) (\dm ^{2} - \sm ^{2})\cos 2\theta}
{\textstyle 2 \um\cm\dm\sm}
%$}}
\label{parlib}
\end{equation}

\noi This last step gives the impression one goes round in a
circle; it does of course not increase the predictive power of the
reasoning, but it allows us to shed a new light on the result. One
might indeed think of $\lambda$ as a physical quantity (since it is
built from physical ones) which is {\it not} renormalized, i.e. which
is not useful for absorbing divergences. Or else, the one-loop $\lambda$
differs from the tree-level one by a finite quantity. In other words,
the Sandard Model with two quark generations and one Higgs doublet seems to
be vast enough to allow one of its apparently free parameters
not to absorb any divergent one-loop radiative correction (we show in
the Appendix that $\delta \,\lambda = 0$ as expected by construction,). This
means that
the one-loop $\beta$-function of $\lambda$ vanishes, i.e. at the one-loop
level, $\lambda$ does not run; $\lambda$ is {\it stable}.

Stating that several natural relations potentially exist in a
one-Higgs-doublet model apparently contradicts previous results obtained in the
context of family symmetries \cite{Wyler:1979, Barbieri:1978,
  Gatto:1979a, Gatto:1979b}. But since we do not appeal to such kind
of symmetries, we do not expect our result to respect the conclusions
derived in their context. Namely, the potential natural relations
(\ref{sol}) cannot be associated with the presence of any extra
horizontal symmetry\footnote{According to \cite{Barbieri:1978}, in a
  one-Higgs-doublet model, the Cabibbo angle which is determined by a
  family symmetry can only take the values $0$ or $\pi/2$,
  yielding $\cos 2\theta = 1$ or $\cos 2\theta = -1$ respectively. In our
  approach, (\ref{parlib}) then reads $\lambda
= (\um ^{2}\dm ^{2} + \cm ^{2}\sm ^{2})/(\um\cm\dm\sm)$ or $\lambda
= (\cm ^{2}\dm ^{2} + \um ^{2}\sm ^{2})/(\um\cm\dm\sm)$, and there
still is an opportunity for relating the mass
ratios among each other. However, again, this potential natural
relation is not to be regarded as the consequence of the presence of
any family symmetry.}. One should examine the validity of the
results at
the n-loop level; then look for the reason of the existence of a {\it
  stable} parameter inside the SM; and finally look
for some possible ``determination principle'' of it outside or beyond the
SM. 

The analysis we have conducted here can be applied to the leptons,
provided that the neutrinos are massive, their mass being of the Dirac
type exclusively. The result, since it depends on the sole structure of
the Yukawa sector, is strictly identical. 

The extension of the present calculation to a three-quark-generation
SM seems to be doomed to failure because of the complexity of the
one-loop radiative corrections to the Cabibbo-Kobayashi-Maskawa parameters
\cite{Cabibbo:1963, Kobayashi:1973}. Those
corrections are indeed too cumbersome to be manipulated and introduced
into a solvable partial differential equations system. We will however
expound, in a forthcoming paper, an alternative approach to
derive similar results in a $n$-quark-generation SM.

\section*{Acknowledgments}
This work is supported by the {\it Fonds pour la Formation {\`a} la
  Recherche dans l'Industrie et dans l'Agriculture} (FRIA).  
We would like to thank J.-M. G{\'e}rard and J. Weyers for a critical
reading of the manuscript.

\section*{Appendix}

\appendix

\section{Self-energies in the Yukawa sector}

After spontaneous breakdown of the symmetry, the Yukawa sector
Lagrangian for quarks reads:
\begin{equation*}
\begin{matrix}
\mathcal{L}_{Y} &
= & 
\bar u_{L}  \Gamma_{d} d_{R} \phi^{+} & 
+ & 
\bar d_{L}  \Gamma_{d} d_{R} \phi^{0} & 
+ & 
\bar d_{R}  \Gamma_{d}^{\dag} u_{L} \phi^{-} & 
+ & 
\bar d_{R}  \Gamma_{d}^{\dag} d_{L} \phi^{0\,\star} 
\\
& 
+ & 
\bar u_{L}  \Gamma_{u} u_{R} \phi^{0\,\star} & 
- & 
\bar d_{L}  \Gamma_{u} u_{R} \phi^{-} & 
+ & 
\bar u_{R}  \Gamma_{d}^{\dag} u_{L} \phi^{0} & 
- & 
\bar u_{R}  \Gamma_{u}^{\dag} d_{L} \phi^{+} 
\\
& 
+ & 
\bar d_{L}  M_{d} d_{R} & 
+ & 
\bar u_{L}  M_{u} u_{R} & 
+ & 
\bar d_{R}  M_{d}^{\dag} d_{L} & 
+ & 
\bar u_{R}  M_{d}^{\dag} u_{L} 
\end{matrix}
\end{equation*}

\noi The Lagrangian fields and parameters are renormalized: 
\begin{equation*}
\begin{matrix}
u_{L,R} & 
\longmapsto & 
u'_{L,R} & 
= & 
(Z^{u}_{L,R})^{-\frac{1}{2}}u_{L,R} \\ 
d_{L,R} & 
\longmapsto & 
d'_{L,R} & 
= & 
(Z^{d}_{L,R})^{-\frac{1}{2}}d_{L,R} \\ 
M_{u} &
\longmapsto & 
M_{u}' & 
= & 
(Z^{u}_{L})^{\frac{1}{2}}
(Z_{M_{u}})^{-1}
M_{u}
(Z^{u}_{R})^{\frac{1}{2}}\\
M_{u} &
\longmapsto & 
M_{d}' & 
= & 
(Z^{d}_{L})^{\frac{1}{2}}
(Z_{M_{d}})^{-1}
M_{d}
(Z^{d}_{R})^{\frac{1}{2}}\\
\end{matrix}
\end{equation*}

\noi The one-loop calculation of the fermion
self-energies leads to (the first
term corresponding to the neutral current intervention ; the second, to
the charged current intervention -- see figure 1): 
\begin{equation}
\begin{matrix}
Z^{u}_{L}
=
1 - \frac{\epsilon}{2} [\Gamma_{u} \Gamma_{u}^{\dag} + \Gamma_{d} \Gamma_{d}^{\dag} ] & & &
Z^{d}_{L}
=
1 - \frac{\epsilon}{2} [\Gamma_{d} \Gamma_{d}^{\dag} + \Gamma_{u} \Gamma_{u}^{\dag} ] \\
Z^{u}_{R}
=
1 - \frac{\epsilon}{2} [\Gamma_{u} \Gamma_{u}^{\dag} + \Gamma_{u} \Gamma_{u}^{\dag} ] & & &
Z^{d}_{R}
=
1 - \frac{\epsilon}{2} [\Gamma_{d} \Gamma_{d}^{\dag} + \Gamma_{d} \Gamma_{d}^{\dag} ]\\
Z_{M_{u}}
=
1 - \epsilon [ 0 + \Gamma_{d} \Gamma_{d}^{\dag}] & & &
Z_{M_{d}}
=
1 - \epsilon [ 0 + \Gamma_{u} \Gamma_{u}^{\dag}]
\end{matrix}
\label{conren}
\end{equation}

\noi with $\epsilon = \left (\frac{1}{4\pi^{2}}\ln
  \frac{\Lambda^{2}}{\mu^{2}}\right )$ where
$\Lambda$ is a cut-off and $\mu$ an arbitrary energy scale (one checks that
$Z^{u}_{L} = Z^{d}_{L}$ as expected). Those results
are true only up to a term proportional to the identity in the flavour
space, which would take into account the electromagnetic, weak
transversal and strong contributions, as well as the tadpole ones. 
But since this term would factor out in the final result, which is
supposed to 
involve exclusively mixing parameters and {\it up}- or {\it down}-type mass
ratios, we chose not to write it down. The finite parts of the
diagrams are omitted. 

From (\ref{conren}), one derives the mass matrices corrections in
the weak base
\begin{align*}
M'_{u} & = M_{u} 
+ \epsilon \;[ M_{d}M_{d}^{\dag}M_{u} - M_{u}M_{u}^{\dag} M_{u} ]
\nonumber\\
M'_{d} & = M_{d} 
+ \epsilon \;[ M_{u}M_{u}^{\dag}M_{d} - M_{d}M_{d}^{\dag} M_{d} ]
\end{align*}

\noi and in the physical base
\begin{align*}
U_{L}^{\dag}M'_{u}U_{R} & = D_{u} 
+ \epsilon \;[K D_{d}^{2}K^{\dag}D_{u} - D_{u}^{3} ]
\nonumber\\
V_{L}^{\dag}M'_{d}V_{R} & = D_{d} 
+ \epsilon \;[K^{\dag} D_{u}^{2}KD_{d} - D_{d}^{3} ]
\end{align*}

\noi where we have absorbed a $\frac{3}{4} \frac{2}{v^{2}}$ factor in $\epsilon$,
$v$ being the scalar VEV, 
%$\epsilon \mapsto  \frac{3}{4} \frac{2}{v^{2}} \epsilon = \frac{3}{4} \frac{2}{v^{2}}\left (\frac{1}{4\pi^{2}}\ln
%  \frac{\Lambda^{2}}{\mu^{2}} \right )$ 
and where $D_{u}$ and $D_{d}$ are
the tree-level diagonal mass matrices while $K$ is the tree-level
Cabibbo-Kobayashi-Maskawa matrix.  Let us repeat that those expressions
do not include the finite parts of the radiative correction, and that
they account for the sole (neutral and charged) scalar exchanges
in the fermion self-energies. 

One can rewrite those last expressions as follows
\begin{align*}
M''_{u}& = D_{u} 
+ \epsilon _{u}
\nonumber\\
M''_{d}& = D_{d} 
+ \epsilon _{d}
\end{align*}

\noi and proceed to the diagonalization of $M''_{u}$ and $M''_{d}$,
i.e. 
\begin{align*}
U_{L}^{\prime \dag}M''_{u}U'_{R} & = D'_{u} 
\nonumber\\
V_{L}^{\prime \dag}M''_{d}V'_{R} & = D'_{d} 
\end{align*}

\noi where
\begin{equation*}
D'_{u}=
\begin{pmatrix}
m_{u}+\epsilon_{u\, 11} &
\\
&
m_{c}+\epsilon_{u\, 22}
\end{pmatrix}
\qquad
U'_{L}=
\begin{pmatrix}
1 & \theta _{u} \\
-\theta _{u} & 1
\end{pmatrix}
\end{equation*}
\begin{equation*}
D'_{d}=
\begin{pmatrix}
m_{d}+\epsilon_{d\, 11} &
\\
&
m_{s}+\epsilon_{d\, 22}
\end{pmatrix}
\qquad
V'_{L}=
\begin{pmatrix}
1 & \theta _{d}\\
-\theta _{d} & 1
\end{pmatrix}
\end{equation*}

\noi with 
\begin{equation*}
\theta _{u}= \frac 
{m_{u}\epsilon_{u\, 21} + m_{c}\epsilon_{u\, 12}} 
{m_{c}^{2}-m_{u}^{2}}
\qquad\mbox{and}\qquad
\theta _{d}= \frac 
{m_{d}\epsilon_{u\, 21} + m_{s}\epsilon_{u\, 12}} 
{m_{s}^{2}-m_{d}^{2}}
\end{equation*}

\noi The one-loop mixing matrix is defined by 
\begin{equation*}
K' = U^{\prime \dagger}_{L} K V'_{L}
\end{equation*}

\noi so that the one-loop mixing angle reads
\begin{equation*}
\theta' = \theta - \theta _{u} + \theta _{d}
\label{varckm}
\end{equation*}

\noi Inserting the value of $\epsilon_{u}$ and $\epsilon_{d}$ in $\theta'$,
$D'_{u}$ and $D'_{d}$, leads to the equations (\ref{varpara}).

\section{Divergent radiative corrections to $\lambda$}

\begin{align*}
\delta \lambda 
& = 
\lambda _{\shortmid \theta}\delta\theta + 
\lambda _{\shortmid \ru}\delta\ru + 
\lambda _{\shortmid \rd}\delta\rd 
\nonumber\\
& = 
- 2 \sin 2 \theta \frac {(1-\ru ^{2})(1-\rd ^{2})}{2\ru\rd}\delta\theta
\nonumber\\
& \qquad + 
\frac{-\rd(1-\ru ^{2})(1+\rd ^{2}) - \rd(1+\ru ^{2})(1-\rd ^{2})\cos 2\theta}
{2 \ru ^{2}\rd ^{2}}\delta\ru
\nonumber\\
& \qquad + 
\frac{-\ru(1+\ru ^{2})(1-\rd ^{2}) - \ru(1-\ru ^{2})(1+\rd ^{2})\cos 2\theta}
{2 \ru ^{2}\rd ^{2}}\delta\rd
\nonumber\\
& =
- \epsilon \sin ^{2}2\theta
\left [
\frac {(1+\ru ^{2})(1-\rd ^{2})}{2\ru\rd}
(\dm ^{2} - \sm ^{2}) 
+
\frac {(1-\ru ^{2})(1+\rd ^{2})}{2\ru\rd}
(\um ^{2} - \cm ^{2}) 
\right ]
\nonumber\\
& \qquad + 
\epsilon \sin ^{2}2\theta
\frac {(1+\ru ^{2})(1-\rd ^{2})}{2\ru\rd}
(\dm ^{2} - \sm ^{2}) 
\nonumber\\
& \qquad + 
\epsilon \sin ^{2}2\theta
\frac {(1-\ru ^{2})(1+\rd ^{2})}{2\ru\rd}
(\um ^{2} - \cm ^{2}) 
\nonumber\\
& = 0
\nonumber\\
\end{align*}

\bibliographystyle{unsrt}
\bibliography{Bibli}

\begin{thebibliography}{10}

\bibitem{Weinberg:1974}
S.~Weinberg.
\newblock {\em Rev. Mod. Phys.}, 46:255--277, 1974.

\bibitem{Weinberg:1977}
S.~Weinberg.
\newblock {\em Trans. New York Acad. Sci.}, 38:185--201, 1977.

\bibitem{Fritzsch:1977}
H.~Fritzsch.
\newblock {\em Phys. Lett.}, B70:436, 1977.

\bibitem{Harari:1978}
H.~Harari, H.~Haut, and J.~Weyers.
\newblock {\em Phys. Lett.}, B78:459, 1978.

\bibitem{Pakvasa:1978}
S.~Pakvasa and H.~Sugawara.
\newblock {\em Phys. Lett.}, B73:61, 1978.

\bibitem{Fritzsch:1990}
H.~Fritzsch and J.~Plankl.
\newblock {\em Phys. Lett.}, B237:451, 1990.

\bibitem{Wyler:1979}
D.~Wyler.
\newblock {\em Phys. Rev.}, D19:330, 1979.

\bibitem{Barbieri:1978}
R.~Barbieri, R.~Gatto, and F.~Strocchi.
\newblock {\em Phys. Lett.}, B74:344--346, 1978.

\bibitem{Gatto:1979a}
R.~Gatto, G.~Morchio, and F.~Strocchi.
\newblock {\em Phys. Lett.}, B83:348--350, 1979.

\bibitem{Gatto:1979b}
R.~Gatto, G.~Morchio, and F.~Strocchi.
\newblock {\em Phys. Lett.}, B80:265, 1979.

\bibitem{Cabibbo:1963}
N.~Cabibbo.
\newblock {\em Phys. Rev. Lett.}, 10:531--532, 1963.

\bibitem{Kobayashi:1973}
M.~Kobayashi and T.~Maskawa.
\newblock {\em Prog. Theor. Phys.}, 49:652, 1973.

\end{thebibliography}

\vspace{1cm}

\hspace{-8mm}
\includegraphics[width=127mm,keepaspectratio]{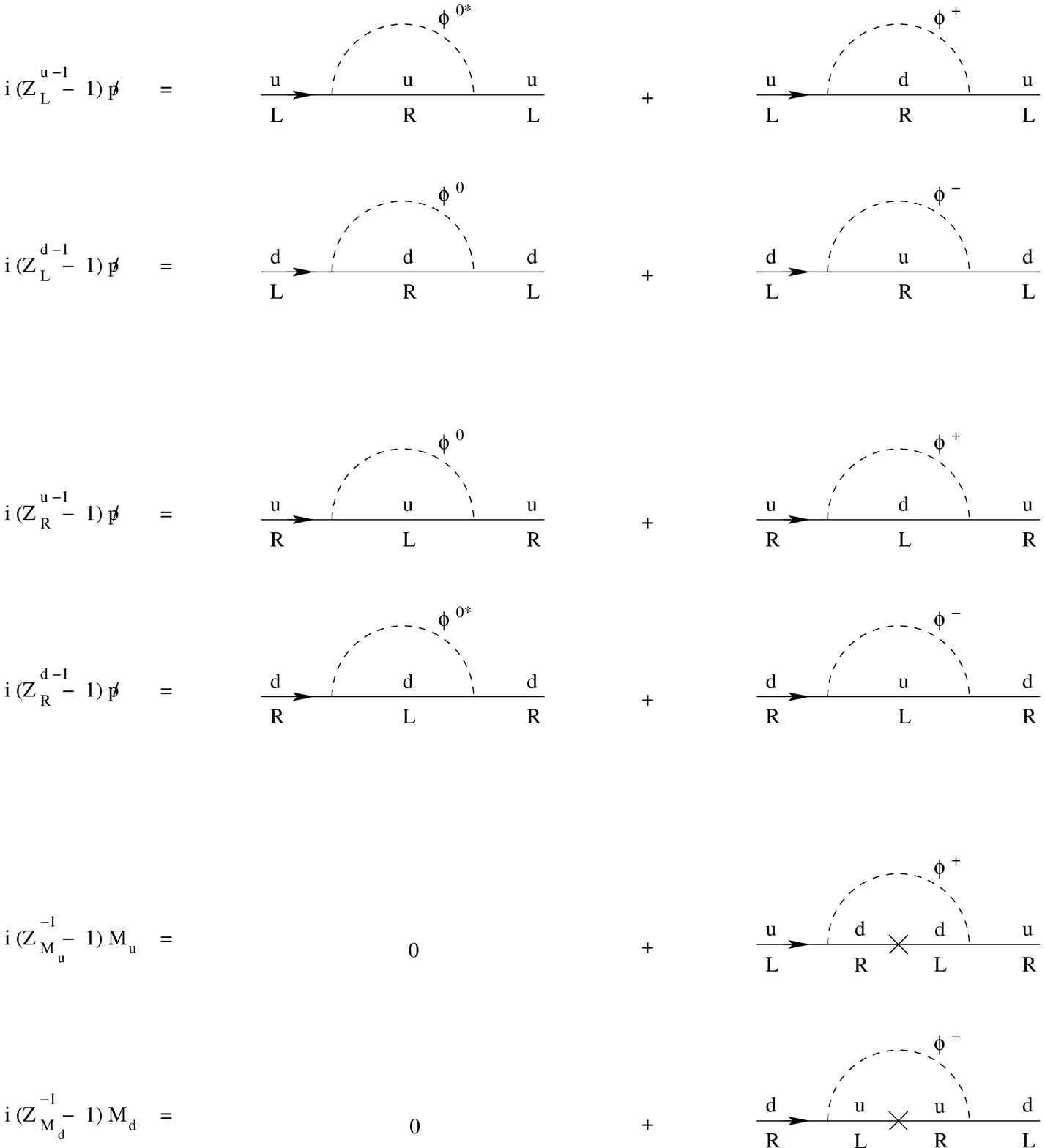}

\vspace{12mm}

\noi Fig. 1: relevant divergent diagrams involved in the calculation
of the renormalization constants.

\end{document}